\documentstyle[11pt,aaspp4]{article}
\input epsf

\def\ep{\epsilon}
\def\ren{R_{M}}

\def\ni{\noindent}

\def\bar{\overline}

\def\OB{\overline{\bf B}}
\def\emf{\overline{\mbox{${\cal E}$}} {}}
\def\emfb{\overline{\mbox{\boldmath ${\cal E}$}} {}}
\def\bbE{\bar {\bf E}}

\def\beq{\begin{equation}}
\def\ee{\end{equation}}
\def\lsim{\mathrel{\rlap{\lower4pt\hbox{\hskip1pt$\sim$}}
    \raise1pt\hbox{$<$}}}
\def\gsim{\mathrel{\rlap{\lower4pt\hbox{\hskip1pt$\sim$}}
    \raise1pt\hbox{$>$}}}
\def\bfE{{\bf E}}
\def\bfJ{{\bf J}}

\def\bfA{{\bf A}}
\def\bfa{{\bf a}}

\def\bfB{{\bf B}}
\def\bbJ{\bar {\bf J}}

\def\ts{\times}
\def\lb{\langle}
\def\rb{\rangle}
\def\curl{\nabla {\ts}}

\def\bfv{{\bf v}}

\def\bfj{{\bf j}}

\def\bfb{{\bf b}}

\def\bfB{{\bf B}}

\def\bbB{\overline {\bf B}}
\def\bbA{\overline {\bf A}}

\def\div{\nabla\cdot}

\begin{document}

\setcounter{equation}{0}

\centerline{\large\bf Dynamical magnetic relaxation: A nonlinear 
magnetically}
\centerline{\large\bf driven dynamo}

\medskip

\author{Eric G. Blackman \altaffilmark{1} and George B. Field\altaffilmark{2}}
\affil{1. Department of Physics \&Astronomy and Laboratory for
Laser Energetics, University of Rochester, Rochester NY 14627}
\affil{2. Harvard-Smithsonian Center for Astrophysics, 60 Garden St.
Cambridge MA, 02138}

\bigskip
\centerline {(in press,  Physics of Plasmas)} 

\begin{abstract} 

\noindent 
A non-linear, time-dependent,  
magnetically driven dynamo theory 
which shows how magnetically dominated configurations can relax
to become  helical on the largest scale 
available is presented.  Coupled time-dependent 
differential equations for 
large scale magnetic helicity, small scale magnetic helicity,  
velocity, and the electromotive force are solved. 
The magnetic helicity on small scales relaxes to
 drive significant large scale helical field growth on dynamical 
(Alfv\'en crossing) time scales, independent of 
the magnitude of finite microphysical 
transport coefficients, after which the growing kinetic helicity 
slows the growth to a viscously limited pace.
This magnetically driven dynamo complements  the nonlinear 
kinetic helicity driven dynamo; for the latter, 
the growing magnetic helicity fluctuations suppress, 
rather than drive, large scale magnetic helicity growth.
A unified set of equations accommodates both types of dynamos.
\end{abstract}

\centerline{PACS codes: 52.30.Cv, 95.30.Qd, 
52.65Kj, 52.55 -s, 96.60.Hv, 98.62Mw}


\bigskip


\section*{I. Introduction}



The concept of dynamo means different things
to different communities: In the astrophysical 
context, dynamo typically refers to  the amplification
of initially weak magnetic fields in a turbulent
flow, but comes in two basic varieties: 
(1) small scale nonhelical (direct) dynamos
in which magnetic energy is amplified by 
random walk field line stretching on spatial or temporal scales less than
or equal to  that of the turbulent forcing 
[1-4],
and (2) large scale helical (inverse) dynamos 
[5-12],
in which helical turbulence produces magnetic energy on scales
larger than that of the input turbulent forcing.
The latter is the type of dynamo 
needed to explain the large scale field and solar cycle of the sun.
It thrives on a non-dissipative term in Ohm's
law called the turbulent electromotive force (EMF), 
$\emfb$, 
which equals the correlated average  (time, space or ensemble)
of the cross product of fluctuating velocity $\bfv$ 
and fluctuating magnetic field $\bfb$,
and is sustained by helical turbulence.

In fusion devices, the plasma is already magnetically dominated.
Here the dynamo acts to convert magnetic flux from
 toroidal to poloidal (or vice versa), increase the scale
of the field, and sustain the strong magnetic flux against
microphysical dissipation 
[13-21].
Like the velocity-driven 
astrophysical  helical dynamo, the magnetically dominated dynamo thrives on 
having a finite  $\emfb$ term in Ohm's law, 
again resulting from correlated turbulent fluctuations
possessing helicity (more on this later).  
But here the fluctuations are driven by current instabilities
and  magnetic helicity injection rather than kinetic helicity.

Although traditional astrophysical
dynamos are normally studied for systems with initially weak magnetic
fields, there are also important magnetically dominated 
 astrophysical environments: astrophysical coronae. These include
solar and stellar coronae, as well as coronae
above accretion disks [23-25]. 
In coronae, the dynamo can  convert and sustain flux, playing a similar role
to the magnetically dominated dynamo in fusion devices.
Coronal systems are magnetically driven
in the sense that helical magnetic fields are injected
 from the disk or star below, and the subsequent time evolution 
is of interest.  A key question is:  
how does the field open up and 
relax to large scales? This is relevant for the formation of
magnetically driven jets and outflows from disks and stars.

It is useful to recognize that a magnetically dominated dynamo 
can also be thought of as dynamical magnetic relaxation. 
Magnetic relaxation describes  the process 
by which magnetic structures in magnetically dominated environments 
evolve to their equilibrium states. 
The fully relaxed end state is the Taylor state
 [\cite{taylor86}], 
determined by minimizing the magnetic energy subject
to the constraint that magnetic helicity is conserved.
The result is a force-free helical configuration
with the scale of the field reaching the largest scale available,  
subject to boundary conditions.
But Taylor's theory by itself is not a dynamical theory 
since it does not provide a time-dependent  description of how 
the large scale magnetic helicity evolves.  
For that, a fully time dependent dynamo theory is needed.
That is our goal herein.

%

In Ref. [\cite{bf02}],  a set of nonlinear dynamo equations for the  
growth of the large scale magnetic field  
was derived and solved when helical turbulent velocity forcing is  applied. 
The nonlinear backreaction due to the build up of small
scale magnetic helicity was shown to ultimately quench the large
scale field growth, in quantitative agreement  with numerical
simulations [\cite{b2001,maronblackman}].   In the present paper, we study the 
complementary problem of large scale field growth  
from a dynamo driven by small scale magnetic helicity fluctuations.
As we will show, the complementarity arises because
the kinetic helicity  becomes the quenching agent rather than the
driving agent for the magnetic helicity driven dynamo.
A  novel feature of the present approach
is that we include a fully time-dependent
dynamical equation for the turbulent EMF which
 couples magnetic helicity and kinetic helicity,
dynamically into the time dependent theory.
Previous work on magnetically dominated dynamos 
focused on more specific magnetized configurations 
[13-21]
 in the steady-state in which $\emfb$ is not solved for dynamically.
We find that coupling the time evolution of $\emfb$ 
 into the theory is 
essential for understanding  time-dependent magnetic relaxation.

We solve the dynamical nonlinear dynamo equations for a 
closed or periodic system. This allows us to ignore boundary terms and
will facilitate future 
testing of the basic principles with tractable 3-D numerical simulations.
The theory is subtle enough, with enough new features,
that we wish to present it for  as  simple a system as possible 
without focusing on detailed magnetic configurations and boundary terms.
These will have to be considered in future work. 

In section II we derive the  coupled 
equations to be solved: the time evolution of large
and small scale magnetic helicities, the turbulent EMF,
and the kinetic helicity. In section III we discuss the solutions
and the physical interpretation, and conclude in section IV.


\section*{II. Derivation of the Coupled Dynamical Equations to be Solved}

We will use a simple multi-scale approach 
that has been relatively successful [\cite{bf02}]
in accounting for the non-linear dynamics of recent 3-D MHD
helical dynamo simulations [\cite{b2001,maronblackman}]
(and has been further been 
shown to be reasonably consistent even compared to a theory
in which 2 additional scales
are included in the dynamics [\cite{blackman03}]): 
We write all quantities as the sum of large scale (indicated by overbar) 
and fluctuating (indicated by lower case) contributions.
The vector potential $\bf A$, magnetic field $\bfB=\curl{\bf A}$ 
(written in Alfv\'en units) and normalized current density $\bfJ=\curl \bfB$
then satisfy $\bfA=\bbA+\bfa$, 
$\bfB=\bbB+\bfb$ and $\bfJ=\bbJ+\bfj$ 
respectively, where the overbar represents a local spatial average.
We assume a separation of scales such that the lower case quantities
vary on scale of inverse wavenumber 
$k_2^{-1}$ and the overbarred quantities $\bbB$, $\bbA$ 
vary on scale $k_1^{-1}$ ($0 < k_1 \ll k_2$).
We also define a global spatial average 
taken over the system scale ($\gg k_1^{-1}$), or periodic box.
The global averages  are indicated by brackets $\lb\rb$,
and being global averages, have no spatial dependence in our approach, only
a time dependence.

We define the total average magnetic helicity 
$H^M=\lb\bfA\cdot\bfB\rb$,  and large and small scale averaged magnetic
helicities as $H_1^M\equiv \lb \bbA\cdot\bbB\rb$
and $H_2^M\equiv\lb\bfa\cdot\bfb\rb$, such that $H^M=H_1^M+H_2^M$.  
The induction equation 
\beq
\partial_t {\bf B}= -\curl {\bf E}=
\curl(\bfv \ts {\bf B}) +\lambda\nabla^2{\bf B},
\label{induction}
\ee
(where  $\bfv$ is the fluctuating velocity, $\lambda$ is the magnetic diffusivity) 
and 
$\bf E= -\partial_t {\bf A} -\nabla \phi$,
imply that [\cite{moffatt}]
\beq
\partial_t H^M=\partial_t H_1^M+\partial_t H_2^M
=-2\lb{\bfE}\cdot\bfB\rb=-2\lambda\lb\bfJ\cdot\bfB\rb,
\label{total}
\ee
where $\phi$ is the scalar potential.
The large and small scale integrated magnetic helicity equations  satisfy  
\beq
\partial_tH_1^M=2\lb{\emfb}\cdot\bbB\rb-2\lambda \lb{\bbB\cdot \bbJ}\rb,
\label{h1}
\ee
and
\beq
\partial_tH_2^M=-2\lb{\emfb}\cdot\bbB\rb -2\lambda 
\lb{\bfb\cdot\bfj}\rb,
\label{h2}
\ee
where the turbulent electromotive force $\emfb\equiv \overline{\bfv\ts\bfb}=
-{\bbE} +\lambda \lb\bbJ\cdot\bbB\rb$.
Note that in deriving (\ref{h2}), we have used 
(\ref{total}), (\ref{h1}), and the definition of $\emfb$.


For later use, it is helpful to derive relations
between globally averaged current helicity, globally
averaged magnetic helicity, and globally averaged magnetic
energy. 
For the current helicity we have 
\beq
\lb{\bf J}\cdot {\bf B}\rb 
= -\lb\bfB \cdot \nabla^2 \bfA \rb + 
\lb\bfB\cdot \nabla(\nabla \cdot \bfA)\rb 
=  -\lb\bfB \cdot \nabla^2 \bfA \rb 
\label{cou}
\ee
where the last equality follow from the chain rule,  
$\nabla\cdot \bfB=0$ and the fact that total divergences vanish
when globally averaged. Correspondingly, 
 $|\lb\bfb\cdot\bfj\rb| =| k_2^2H_2^M| \le k_2 \lb\bfb^2\rb$  
and $|\lb\bbB\cdot\bbJ\rb| = |k_1^2H_1^M| \le k_1\lb\bbB^2\rb$.  
The inequalities in the preceding two relations
represent the the realizability condition [\cite{fplm}] 
and the equality holds for maximally
helical (force-free) structures on the respective scales.
(These are gauge independent relations because 
globally averaged divergences vanish. If we had used 
the Coulomb gauge $(\nabla\cdot \bfA=0)$, then (\ref{cou}) and the 
relations above would have followed straight away, 
without having to appeal to vanishing global divergences.) 

The dynamical equation for the turbulent $\emfb$ is 
\beq
\partial_t\emfb=\overline{\partial_t
\bfv\ts\bfb}+\overline{\bfv\ts\partial_t\bfb}.
\label{timed}
\ee
Therefore,  we 
need equations for $\partial_t\bfb$ and $\partial_t\bfv$.  
Assuming that $\div\bfv=0$, subtracting the local mean of 
(\ref{induction}) from  (\ref{induction}) 
we have
\begin{equation}
\begin{array}{r}\partial_{t} \bfb = \curl(\bfv\ts\OB) + 
\curl(\bfv\ts\bfb) 
-\curl(\overline {\bfv\ts\bfb}) +
\lambda\nabla^{2}\bfb.
\end{array}
\label{c3}
\ee
Similarly, in the absence of any mean velocity, the equation
for fluctuating velocity becomes 
\begin{equation}
\begin{array}{r}
\partial_{t} {\bfv} = \bfv \ts {\vec \omega}- \overline{\bfv \ts {\bf \vec \omega}}
-\nabla (p + \bfv^2/2  - \overline{ \bfv^2}/2)  
+ {\bf j}\ts \OB + {\bbJ}\ts \bfb
+ \nu\nabla^{2}{\bfv}, 
\end{array}
\label{c2}
\end{equation}
where $\nu$ is the viscosity and $p$ is the fluctuating pressure. 
In deriving  (\ref{c2}) we assume  that 
$\bfj\ts\bfb=0$, from being injected with this property.
For simplicity, we also assume that this is maintained
at all times (this needs to be tested). This assumption 
should not affect the basic conclusions from our study because
only $\bfj\cdot\bfb$ enters the turbulent electromotive force $\emfb$.
In addition, we do not assume that the magnetic force associated with the 
cross-terms between large and small scales are force-free, and thus
we still allow terms such as $\bfj\ts\bbB$ and
$\bbJ\ts\bfb$ to contribute to the small scale magnetic force.


By analogy to Ref. [\cite{bf02}], 
using  (\ref{c3}) and (\ref{c2}) we then have for  (\ref{timed})
\beq
\begin{array}{r}
\partial_t{\emfb}=
{1\over 3}(\overline{\bfb\cdot\curl\bfb}
-\overline{\bfv\cdot\curl\bfv})\bbB 
- {1\over 3}
\overline{\bfv^2}\curl\bbB
+\nu\overline{\nabla^2\bfv\ts\bfb} +\lambda\overline{\bfv\ts\nabla^2\bfb}
+ {\bf T},
\end{array}
\label{timed33}
\ee
where ${T_j}=[\overline{ \bfv\ts\curl(\bfv\ts\bfb)}]_j-\ep_{jqn}\overline{P_{qi}
(\bfv\cdot\nabla v_i) b_n}$
represents  the surviving triple correlations
(by assuming $\bfj\ts\bfb=0$ there
no triple correlations survive with more than one factor of $\bfb$) 
and where 
$P_{qi}\equiv (\delta_{qi}-\nabla^{-2}\nabla_q\nabla_i)$
is the projection operator that arises after taking the divergence
of the incompressible Navier-Stokes equation to eliminate the 
pressure.  In deriving (\ref{timed33}) we took $\bfv$ and $\bfb$ to be statistically isotropic.  It will be of interest
in future work to drop this assumption [\cite{kr,rk}].

For Eqns. (\ref{h1}) and (\ref{h2}), 
we want the component of $\emfb$ parallel to 
$\bbB$. For this we have
\beq
\partial_t\emf_{||}=(\overline{\partial_t\bfv\ts\bfb}
 +\overline{\bfv\ts\partial_t\bfb})\cdot\bbB/|\bbB|+
\overline{\bfv\ts\bfb}\cdot\partial_t(\bbB/|\bbB|).
\label{timedp}
\ee
Using Eqns. (\ref{timed33}) and (\ref{timedp}) we have
\beq
\partial_t\emf_{||}= {\tilde\alpha}{\bbB^2/|\bbB|}
-{\tilde\beta}{\bbB\cdot\curl\bbB}/|\bbB|-{\tilde \zeta}\emf_{||}
\label{2}
\ee
where 
${\tilde\alpha}
=(1/3)(\overline{\bfb\cdot\curl\bfb}-\overline{\bfv\cdot\curl\bfv})$,  
${\tilde\beta} = (1/3)\overline{\bfv^2}$, and  
$\tilde\zeta \equiv fk_2{\overline{\bfv^2}}^{1/2}$
accounts for microphysical dissipation
terms, the last term of (\ref{timedp}), and most importantly, 
${\bf T}$. We will take the constant $f\sim 1$, which follows from estimating the magnitude
of $\bf T$.  
Note that $\tilde \zeta$ and $\tilde \beta$ depend on $\overline{\bfv^2}$ which must be
solved for dynamically.

When $\OB$ is force-free and only one sign of helicity
is initially injected into the volume, 
${\bbB\cdot\bbJ}\simeq \lb{\bbB\cdot\bbJ}\rb$ and
$\OB^2\simeq \lb\OB^2 \rb=k_1|H_1^M|$ [\cite{b2001}].  
(We consider the injection of one sign to illustrate
as simply as possible the tendency for magnetic helicity to inverse cascade.
If both signs are injected with equal magnitude then no net magnetic helicity 
would be injected at all.   Also, the source 
of large scale field growth in the problem we consider comes from the $\emfb$
terms, which grows force-free large scale fields.)
In addition, $\overline{\bfv^2}\simeq \lb\bfv^2\rb$, 
$\overline{\bfv\cdot \curl\bfv
}\simeq \lb\bfv\cdot \curl\bfv\rb$,  and
$\overline{\bfb\cdot \bfj
}\simeq \lb\bfb\cdot \bfj\rb$.
Then  
(\ref{h1}) and (\ref{h2}) become
\beq
\partial_t H_1^M=2\emf_{||}k_1^{1/2}|H_1^M|^{1/2}
-2\lambda k_1^2H_1^M
\label{h3}
\ee
and 
\beq
\partial_t H_2^M=-2\emf_{||}k_1^{1/2}|H_1^M|^{1/2}-2\lambda k_2^2H_2^M.
\label{h4}
\ee
Using $U\equiv\lb\bfv^2\rb$, $H_2^V\equiv\lb\bfv\cdot\curl\bfv\rb$, 
we re-write (\ref{2}) in the two scale approach as
\beq
\partial_t\emf_{||}=
k_1^{1/2}|H_1^M|^{1/2}(k_2^2H_2^M-H_2^V)/3-k_1^{3/2}(H_1^M/(|H_1^M|^{1/2})
{U/3}-{f k_2 U^{1/2}}\emf_{||}.
\label{4time}
\ee
To solve for the evolution of $H_1^M$ and $H_2^M$ using (\ref{4time}),
we also need dynamical equations for $U$ and $H_2^V$. 
Our substitution  $\overline{\bfv^2}\simeq \lb\bfv^2\rb$ allows
us to ignore boundary terms, since we have stated that  $\lb\rb$ is taken
over a closed or periodic volume.
Then, from (\ref{c2}), 
\beq
\partial_t U =2(\lb\bfv\ts \bfj)\cdot\bbB\rb -2\lb(\bfv \ts \bfb)\cdot\bbJ\rb
-2\nu\lb(\nabla \bfv)^2\rb
\simeq 2(k_2-k_1)\lb\emfb\cdot \bbB\rb-2\nu k_2^2 \lb\bfv ^2\rb,
\label{U}
\ee
and 
\beq
\partial_t H_2^V= 2\bbB\cdot\lb{\vec \omega }\ts\bfb\rb(k_2-k_1)
-2\nu\lb\partial_iv_j\partial_i\omega_j\rb
\simeq 2k_2(k_2-k_1)\lb\emfb\cdot \bbB\rb-2\nu\lb\partial_iv_j\partial_i\omega_j\rb,
\label{h2v}
\ee
where we consider the case that  $H_1^M,H_2^M > 0$, took advantage of 
$\bfj\ts \bfb =\bbJ\ts\bbB=0$, and 
again dropped surface terms.
We also used $\partial_t \lb\bfv\cdot\bfb\rb = - k_2^2(\lambda+\nu)
\lb\bfv\cdot\bfb\rb$ (showing that $\lb\bfv\cdot\bfb\rb$ decays)
to write $\lb{\vec \omega}\ts \bfb \rb_q = \lb v_s \partial_q b_s\rb
=\lb \bfv \ts \bfj\rb_q$ in deriving the penultimate term of (\ref{h2v}).

The equations to be solved are thus 
(\ref{h3}), (\ref{h4}), (\ref{4time}), (\ref{U}) and (\ref{h2v})
after  converting them into dimensionless form.
We  write $H_2^M$ at $t=0$ as  
$H_2^M(0)$ and 
define the dimensionless quantities
$h_1\equiv H_1^M/H_2^M(0)$, $h_2\equiv H_2^M/H_2^M(0)$,
$\ren\equiv \sqrt{H_2^M(0)}/\lambda k_2^{1/2}$, 
$R_V\equiv \sqrt{H_2^M(0)}/\nu k_2^{1/2}$, 
$\tau\equiv t k_2^{3/2} \sqrt{H_2^M(0)}$, $Q=-\emf_{||}/k_2H_2^M(0)$, 
$\epsilon=U/{k_2 H_2^M(0)}$, and $h_v\equiv H_2^V/k_2^2 H_2^M(0)$. 
Using these in (\ref{h3}), (\ref{h4}), (\ref{4time}), (\ref{U})
and (\ref{h2v}) respectively gives 
\beq
\partial_\tau h_1=-2Qh_1^{1/2}(k_1/k_2)^{1/2}-2h_1(k_1/k_2)^2/\ren,
\label{7}
\ee
\beq
\partial_\tau h_2=2Qh_1^{1/2}(k_1/k_2)^{1/2}-2h_2/\ren,
\label{8}
\ee
\beq
\partial_\tau Q= -\left(k_1/k_2 \right)^{1/2}h_1^{1/2}(1/3) (h_2-h_v)
+(k_1/k_2)^{3/2}h_1^{1/2}\epsilon/3-\epsilon^{1/2}fQ,
\label{6}
\ee
\beq
\partial_\tau\epsilon= -
2(1 - k_1/k_2)(k_1/k_2)^{1/2}Q h_1^{1/2}
-2 \epsilon /R_{V},
\label{epsilon}
\ee
and
\beq
\partial_\tau h_v=-2(1-k_1/k_2)(k_1/k_2)^{1/2}Qh_1^{1/2}-2h_v/R_V
\label{hv2}
\ee
(In the simple treatment above
we have not coupled in an equation for the non-helical
part of the magnetic energy. This would arise for example,
from a bulk large scale shear, which we also do not include.
A varying fraction of non-helical magnetic energy 
on the large scale would add a time dependent coefficient $\ge 1$ to $Q$ 
in (\ref{7}) [\cite{bb03}].)

Below we consider two cases C1 and C2: For C1, $h_2$ takes on an initial
value $h_2(0)=1$, and then evolves according to (\ref{8})
while for C2, $h_2=1$ is fixed for all times.
Case C1 corresponds to free relaxation (simulated in Ref. [\cite{stribling}]) 
and case C2 corresponds to driven relaxation. 
The latter represents injecting $h_2$ into the system, 
for example  by a potential difference [\cite{bellan00,jiprager02}] 
which drives instabilities that keep $h_2$ steady.  
In both cases $\ep$ and $h_v$ satisfy the same equation, 
so $h_v=\epsilon$ for all times when both are initially small.


\section*{III. Discussion of Solutions and Physical Interpretation}

Solutions for case C1 with $f=1$ and different  $R_M=R_V$ 
are shown in Fig. 1. 
We see that $h_1$ grows and $h_2$ depletes in accordance
with magnetic helicity conservation, while at the same time $\epsilon=h_v$ 
grows. This prevents $h_1$ from equaling
$h_2(0)=1$ due to the backreaction from $h_v$ in (\ref{6}).
Instead, the maximum $h_1$ occurs at $h_1\sim h_2 \sim 1/2$,
after which resistive terms take over and all quantities decay.
If $h_v$ did not grow, then $h_1$ would reach $h_1\sim h_2(0)=1$. 

Solutions for case C2 with $f=1$ are shown in Fig 2.
In Fig. 2a and 2b, a significant kinematic regime of duration 
$\tau_{kin}$ in which growth 
of $h_1$ is independent of $R_M$ and $R_V$ is evident.
The quantity $\tau_{kin}$ can be estimated by 
ignoring the $R_M$ and $R_V$ terms and replacing the time
derivatives in (\ref{7}), (\ref{6}), (\ref{epsilon}), and 
(\ref{hv2}) by multiplication by $1/\tau_{kin}$.
The end of the kinematic phase occurs when $h_v$ grows large enough
to deplete the second term in (\ref{6}) 
such that the third term in (\ref{6}) becomes important.
This occurs when $h_v=\ep=(1+k_1/k_2)^{-1}$.
Using this, and solving for $\tau_{kin}$ 
from the approximated versions of (\ref{7}), (\ref{6}) and (\ref{epsilon}) 
gives 
\beq
\tau_{kin}\simeq (3f/2)(k_2/k_1)^2(1+k_1/k_2)^{1/2},
\label{tkin}
\ee
which is $\sim 40$  for $k_2/k_1=5$ and $f=1$.
Solving for $h_1$ at $t_{kin}$, then gives 
$h_1\simeq (1-k_1^2/k_2^2)^{-1}\sim 1$. Thus $h_1$
grows kinematically up to a value  $\sim h_2=1$. 

After $\tau_{kin}$, $R_M$ and $R_V$ become important.
The ratio determines the saturation
value of $h_1$, while $R_V$
determines the rate of approach to the saturation.
That $R_V$ rather than $R_M$ determines this rate
results because the build up of $h_v$ 
suppresses growth of $h_1$  and $R_V$ appears in (\ref{hv2}) 
in the diffusion term. 
We can estimate the final saturation values expected
for case C2 by setting the right sides of 
(\ref{7}), (\ref{6}), (\ref{epsilon}), and 
(\ref{hv2}) equal to zero. The
solution is 
$\epsilon=h_v=(1+k_1/k_2)^{-1}$, 
$Q\simeq -(k_1/k_2)^{1/2}(1-k_1^2/k_2^2)^{-1/2}
(R_V R_M)^{-1/2}\ll 1$, and $h_1=(R_M/R_V)(k_2/k_1)^2(1-k_1^2/k_2^2)^{-1}\simeq (R_M/R_V)(k_2/k_1)^2$. The  saturation value of $h_1$ 
can be seen to depend on $R_V/R_M$ in Fig 2c.

For both cases C1 and C2 we have also solved the dynamical 
equations for $f=1/10$ (C2 solution plotted in Fig. 2d). 
This corresponds to a closure in which the damping time in the last term of 
(\ref{6}) is $\sim 10$.  
Oscillations appear at early times, similar to what is seen 
in Ref. [\cite{bf02}] for the $h_v$ driven dynamo
when $f<1$ and when the closure is applied to 
passive transport of a scalar [\cite{bf03flu,branfluid}].
Unlike the case of Ref. [\cite{bf02}] however, 
$h_v$ evolves here, so the similar sensitivity to $f$ 
is noteworthy.  Here, for $f < 1$, $Q$ oscillates around zero because 
$h_v$ can grow to be larger than $h_1$. 
The sign of $\partial_\tau Q$ then changes
and eventually $Q$ changes sign as well. For  $f < 1$, 
the last term in (\ref{6}) cannot damp  growth of $Q$ fast enough 
to prevent its sign change.
Eventually, the $R_V$ term  takes over in (\ref{epsilon})
and (\ref{hv2}), damping the growth of $\ep$ 
and the oscillations.  The oscillation amplitudes
provide a prediction for a given $f$ and a diagnostic for the closure.  
Note that the rise to the first peak in $h_1$ corresponds
to the end of the kinematic regime. Eqn. (\ref{tkin}) shows
that $\tau_{kin}$ decreases with $f$, explaining why the first peak
occurs earlier for $f<1$ than the end of the kinematic regime for $f=1$
(compare Fig. 2d with Fig. 2b).  

An important feature of the
system of equations (16-20)
is that they  
represent a unified framework for understanding aspects of BOTH 
kinetic helicity driven helical dynamos and dynamical magnetic
relaxation depending on 
whether the source driving the fluctuations
initially supplies kinetic helicity or magnetic helicity.  
If kinetic helicity is steadily supplied, we have
the nonlinear dynamo of Ref. [\cite{bf02}].
Then $\partial_t h_2 =\partial_t\ep= 0$ and 
$Q$ is initially dominated by the
$h_v$ contribution in (\ref{8}) which drives
growth of the large scale magnetic helicity in (\ref{6}).
The near conservation of magnetic helicity then means
that the small scale magnetic helicity $h_2$
must grow with opposite sign to compensate the growth of $h_1$.
The growth of $h_2$,  in turn, offsets
the kinetic helicity in the first term on the right of 
(\ref{8}), slowing the growth of $h_1$ in (\ref{7}), 
and quenching the system into a steady-state.
In contrast, as shown above, when the driving initially supplies
$h_2$ instead of $h_v$,  $Q$ is initially dominated by $h_2$. 
Then $h_v$ grows, offsets $h_2$ in (\ref{8}) which in turn
quenches the growth of $h_1$ and again drives the system into a steady-state.
In this sense, the  kinetic helicity driven dynamo and the magnetically
helicity driven dynamos are fully complementary, and their non-linear
evolution is explicable by the same set of equations with different
initial conditions. In both the kinetic and magnetic helicity
driven dynamos, the large scale magnetic helicity 
incurs an initially fast growth phase (kinematic regime)
which is independent of the 
value of the finite microphysical diffusivities,
as long as the driving fluctuation scale is far enough above the
diffusive scales.

One important difference between two types of dynamos 
is  that for the kinetic helicity driven
dynamo, the large and small scale magnetic helicities have {\it opposite} signs,
whereas for the magnetically driven dynamo, the large and small
scale magnetic helicities have the {\it same} sign.  This is easy to
understand and highlights the role of  magnetic helicity conservation:
In the kinetic helicity driven dynamo, there is initially
negligible total magnetic helicity, and the kinetic helicity
acts to drive one sign to large scales and the other sign
to small scales. In the magnetic helicity driven case, there
is a net initial total magnetic helicity injection  on small scales 
and the dynamo acts to drive an inverse transfer of magnetic
helicity to large scales.

Coming back to the examples discussed in Sec. I,
note that the magnetically dominated dynamo, or dynamical magnetic relaxation
is most relevant for astrophysical coronae or laboratory devices,
where magnetic helicity is injected into the system and the system evolves
to relax dynamically. The kinetic helicity driven dynamo is most
relevant inside  astrophysical objects, which are not magnetically
dominated. Note however that for a star or accretion disk, {\it both} 
dynamos can actually operate symbiotically: The kinetic helicity 
dynamo produces large scale fields inside the disk or star, 
which then rise to the corona.
There the supplied fields are ``small scale'' with respect to the corona,
and can subsequently dynamically relax to larger scales via a magnetically
driven dynamo.
Rapid generation of the large scale fields in this way can be an important
source of large scale coronal magnetic fields for magnetically mediated
outflows.

Certainly our simple study herein cannot yet describe
the details of astrophysical or laboratory applications in detail,
but we have identified some unifying physical principles of
magnetically dominated dynamos, kinetic helicity driven dynamos,
and dynamical magnetic relaxation which can help guide further work.

\section*{IV. Conclusions}
We have developed a nonlinear dynamical theory of
magnetic relaxation, or equivalently, 
a magnetically driven large scale helical dynamo.
To isolate the basic principles 
and to facilitate comparison with the simplest numerical
simulations, we solved the dynamical magnetic relaxation equations 
in a closed (or periodic) system.
Compared to the $h_v$ driven dynamo of Ref. [\cite{bf02}], 
the role of $h_2$ and $h_v$ are reversed in 
dynamical magnetic relaxation: in an $h_2\ (h_v)$ driven dynamo, 
subsequent growth of $h_v\ (h_2)$ quenches the growth of $h_1$.
Also, $h_1$ grows with the same (opposite) sign
of $h_2$ in an $h_2\ (h_v)$ driven dynamo.
We have found that dynamical magnetic relaxation always involves
an initial rapid transfer of magnetic helicity from $k_2$ 
to $k_1$, independent of $R_M$ and $R_V$, followed by 
a slow, $R_V$ dependent evolution of 
$h_1$. When  
$h_2=1$ is fixed, $h_1$ saturates at 
$h_1\simeq (k_2/k_1)^2R_M/R_V$.  When $h_2(0)=1$ 
and $h_2$ is allowed to evolve according to (\ref{8}), 
$h_1\simeq 1/2\simeq h_2$ at maximum before resistively decaying.


More detailed work is needed to incorporate the principles
we have identified 
to specific laboratory configurations or to magnetized astrophysical 
coronae of stars and accretion disks. 
However, our results lead us to 
predict that  significant magnetic relaxation will always occur on 
dynamical time scales 
($\sim$ Alfv\'en crossing time scales):
The large scale magnetic helicity reaches $ \sim 1/2$ 
the strength of the injection scale helicity during a time 
$\tau_{kin}$, after which 
relaxation should be viscously limited.   The magnetic Prandtl
number and the ratio of large to small scales 
then determine the ultimate saturation values.

The theory  can be fully tested  with  3-D MHD numerical
experiments in a periodic box in which 
magnetic helicity is injected at some relatively large 
wavenumber, say $k_f\sim 5$, with an initially negligible velocity. 
The overall evolution of the magnetic helicity and kinetic helicity
spectra can then be measured as a function of time.  
Future generalizations must incorporate boundary terms,
which can be important in both astrophysical and laboratory
contexts.

\ni 
EB acknowledges DOE grant DE-FG02-00ER54600.


{\bf REFERENCES}

\enumerate

\bibitem{kazanstev}A.P. Kazanstev, Sov. Phys. JETP, {\bf 26} 1031 (1968)

\bibitem{kida91}  
 S. Kida, S. Yanase, J. \& Mizushima,  Phys.  Fluids {\bf3} 457 (1991)

\bibitem{maroncowley}
J. Maron, S. Cowley, J. McWilliams, `''The Nonlinear Turbulent Dynamo,''
in press Astrophys. J. (2004)

\bibitem{haugen} N.E.L. Haugen, 
A. Brandenburg, W. Dobler,   Astrophys. J. Lett., {\bf 597}, L141 (2003)

\bibitem{pfl}
A. Pouquet, U. Frisch,  J. L\'eorat,   J. Fluid Mech., {\bf 77}  321 (1976)

\bibitem{moffatt} H.K. Moffatt, {\sl Magnetic
Field Generation in Electrically Conducting Fluids}, 
(Cambridge University Press, Cambridge, 1978)

\bibitem{parker}  
E.N. Parker, {\it Cosmical Magnetic Fields}, (Oxford: Clarendon
Press, 1979)

\bibitem{krause}   F. Krause \& K.-H. R\"adler, 
{\it Mean-field Magnetohydrodynamics and Dynamo Theory}, 
(Pergamon Press, New York, 1980)

\bibitem{zeldovich83} 
Ya. B. Zeldovich , A.A. Ruzmaikin, \& D.D. Sokoloff, {\sl Magnetic Fields in Astrophysics}, 
(Gordon and Breach, New York, 1983)

\bibitem{b2001}
A. Brandenburg, Astrophys. J., {\bf 550}, 824 (2001)

\bibitem{maronblackman}
J. Maron \& E.G. Blackman, Astrophys. J. Lett. {\bf 566}, L41 (2002). 

\bibitem{bf02} E.G. Blackman \& G.B. Field, 
Phys. Rev. Lett., {\bf 89}, 265007 (2002)


\bibitem{ortolani93} 
S. Ortolani  \& D.D. Schnack, 
{\it Magnetohydrodynamics of Plasma Relaxation}
(World Scientific: Singapore, 1993)



\bibitem{strauss85} 
H.R. Strauss, Phys. Fluids, {\bf 28}, 2786 (1985)

\bibitem{strauss86} 
H.R. Strauss, Phys. Fluids, {\bf 29}, 3008 (1986)

\bibitem{bhattacharjee86} 
A. Bhattacharjee \& E. Hameiri,  Phys. Rev. Lett. 
{\bf 57}, 206 (1986)

\bibitem{holmesetal88} 
J.A. Holmes, B.A. Carreras, P.H. Diamond, \& V.E. Lynch,  
Phys. Fluids, {\bf 31}, 1166 (1988)

\bibitem{gd1}  
A.V. Gruzinov \& P.H. Diamond, Phys. Plasmas, {\bf 2}, 1941 (1995)

\bibitem{by}  A. Bhattacharjee \& Y. Yuan, Astrophys. J.,  
{\bf 449}, 739 (1995)

\bibitem{bellan00}  P.M. Bellan, {\sl Spheromaks}, 
(Imperial College Press, London, 2000)

\bibitem{jiprager02} 
 H. Ji \& S.C. Prager,  {Magnetohydrodynamics}
{\bf 38}, 191  (2002)

\bibitem{taylor86}
J.B. Taylor, Rev. Mod. Phys., {\bf 58}, 741 (1986) 

\bibitem{galeev79} 
A.A. Galeev, R. Rosner, G.S. Vaiana, Astrophys. J., {\bf 229} 318 (1979)

\bibitem{sz}
C.J. Schrijver  \& C. Zwaan, {\it Solar and Stellar Magnetic Activity},
(Cambridge: Cambridge Univ. Press, 2000)

\bibitem{fieldrogers93} 
G.B. Field \& R.D. Rogers, R.D. Astrophys. J., {\bf 403} 94 (1993) 

\bibitem{blackman03}
E.G. Blackman, MNRAS, {\bf 344}, 707  (2003). 

\bibitem{fplm} 
 U. Frisch, A. Pouquet, J. L\'eorat \& A. Mazure,  J. Fluid Mech. 
{\bf
68}, 769 (1975).

\bibitem{kr}
N. Kleeorin \& I. Rogachevskii,  Phys. Rev. E., {\bf 59}, 6724 (1999)

\bibitem{rk} I. Rogachevskii 
\& N. Kleeorin, Phys. Rev. E., {\bf 64}, 56307 (2001)

\bibitem{bb03} E.G. Blackman 
\& A. Brandenburg, Astrophys. J. Lett., {\bf 584} L99  (2003)

\bibitem{stribling} T. Stribling, W.H. Matthaeus, \& S. Ghosh,  
J. Geophys.\ Res., {\bf 99}, 2567 (1994)

\bibitem{bf03flu} E.G. Blackman \& 
G.B. Field, Phys. Fluids, {\bf 15}, L73 (2003)

\bibitem{branfluid}
A. Brandenburg, P. K\"apyl\"a, A. Mohammed, 
``Non-Fickian Diffusion and the tau-approximation from
numerical turbulence,'' astro-ph/0306521,
submitted to Phys. Fluids. (2003).

\eject

\noindent{\bf Figure 1}: 
Solutions 
for $k_1=1$, $f=1$, $k_2=5$, 
(a) $R_M=R_V=200$, case C1: $h_2$ free 
(b) $R_M=R_V=2000$, case C1: $h_2$ free 
The curves in (a) and (b) are identified as follows:
$h_1$ is the thick solid line; $h_2$ is the thin solid line;
$h_v$ is the long dashed line; $-Q$ is the short dashed line.

\noindent {\bf Figure 2}: 
(a) Same as Fig. 1a, but case C2: fixed $h_2=1$.
(b) Same as Fig. 1b, but case C2: fixed $h_2=1$.
(c) $h_1$ in case C2 for late times: solid curve is for $R_M=2000$, $R_V=200$; 
short dashed curve is for $R_M=R_V=200$; long dashed curve is for 
$R_M=R_V=2000$.
(d) Same as (b),  but with $f=1/10$. Notice that 
$-Q$ now oscillates about 0.
The curves in (a) and (b) are identified as follows:
At late times (not shown) the oscillations damp, and the 
solutions are indistinguishable from the $f=1$ case.

{\bf FIGURE 1ab}: 
\bigskip
\bigskip

\vspace{-.1cm} \hbox to \hsize{ \hfill \epsfxsize12.3cm
\epsffile{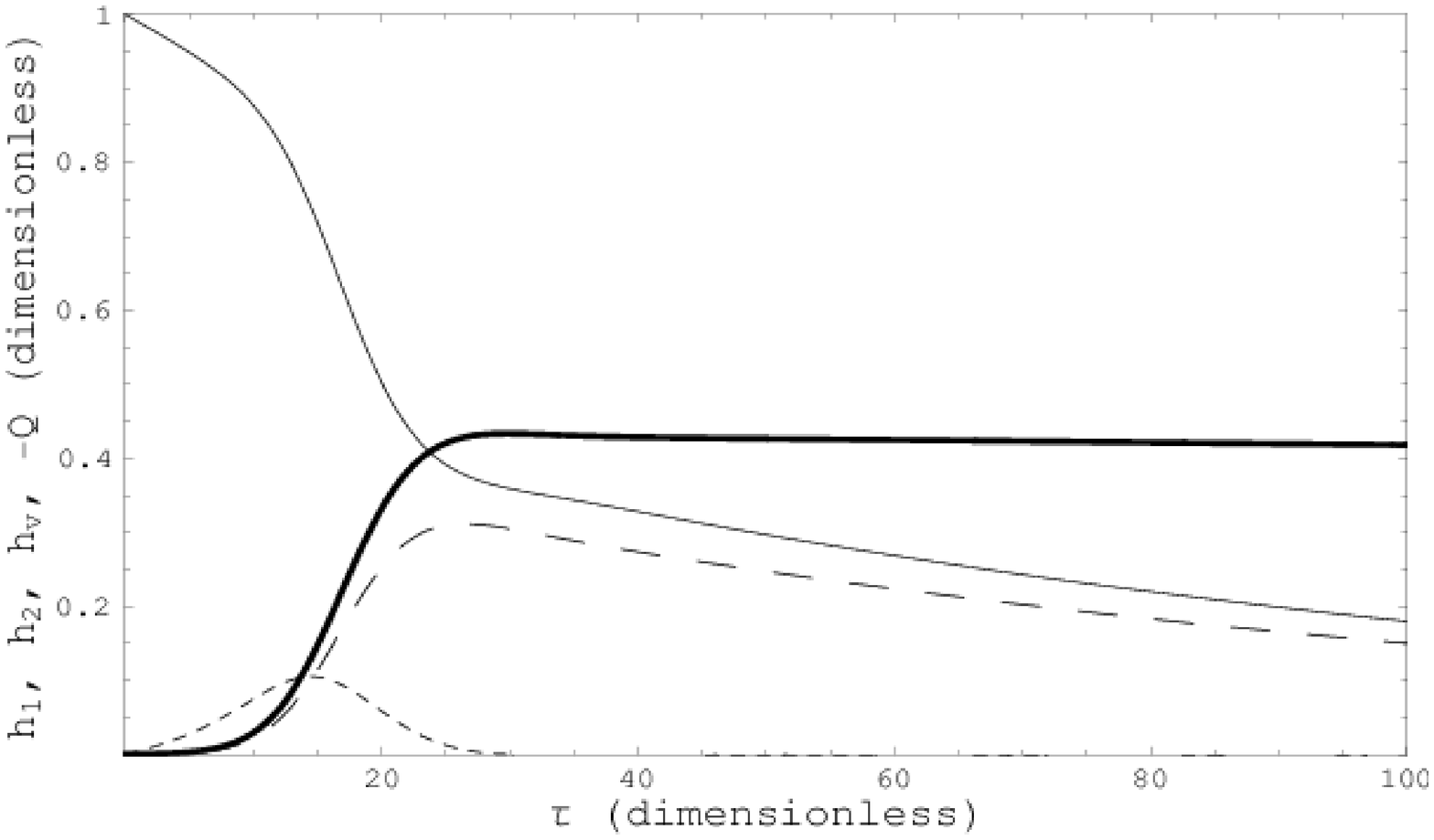} 
\hfill }
\vspace{0cm} \hbox to \hsize{ \hfill \epsfxsize12.3cm \epsffile{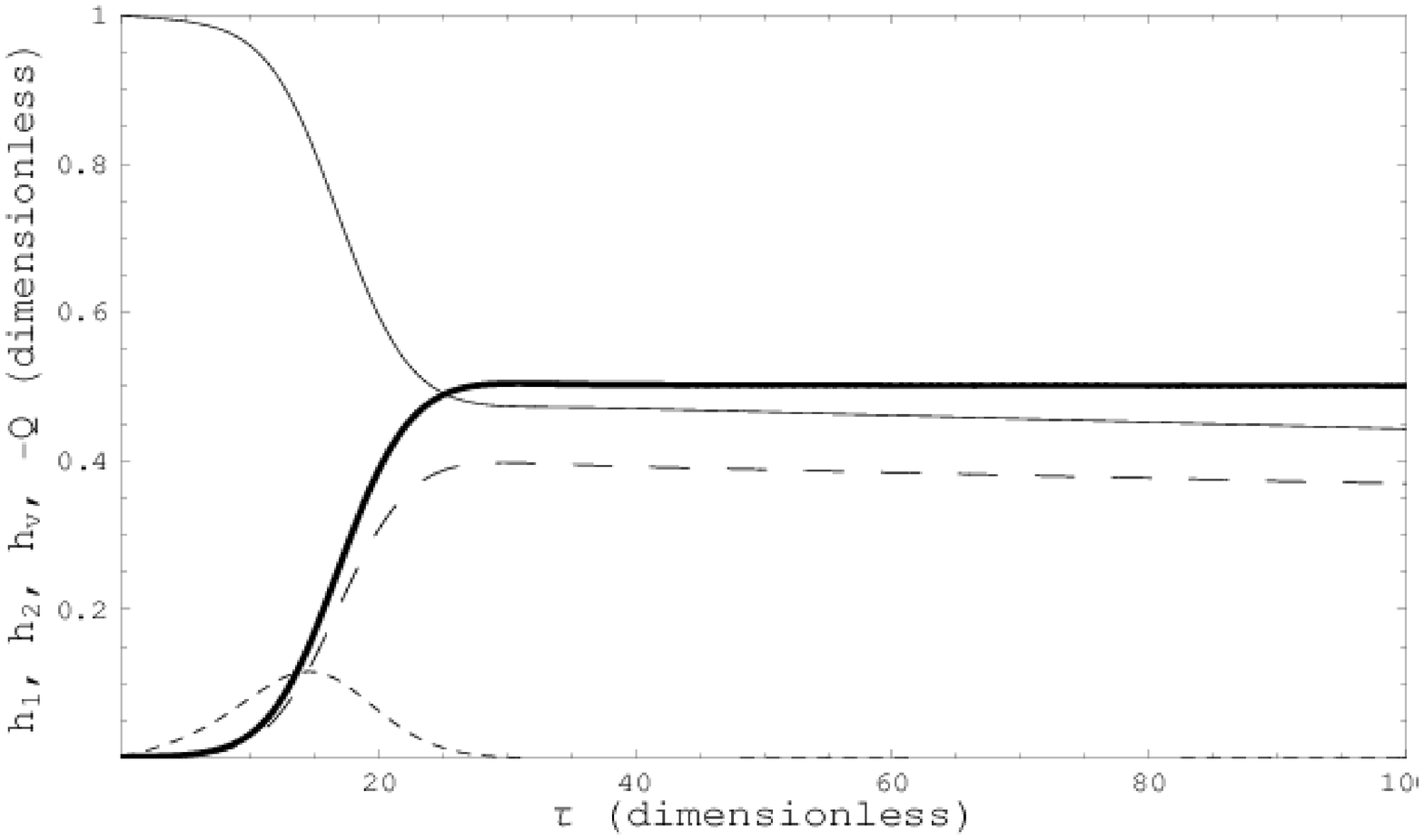} 
\hfill }

\eject
{\bf FIGURE 2ab:} 
\bigskip




\vspace{0cm} \hbox to \hsize{\hfill \epsfxsize14cm \epsffile{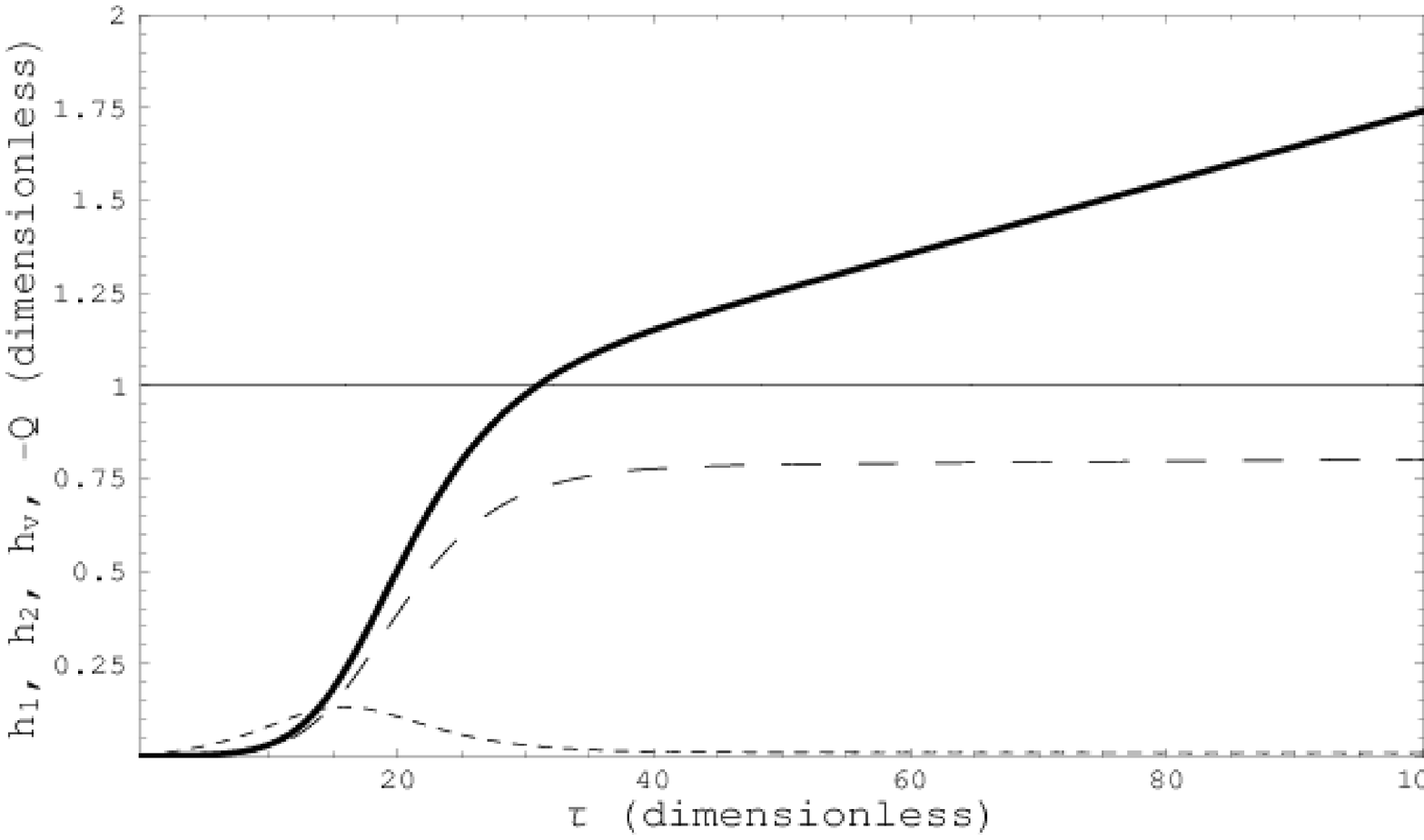}
\hfill}

\vspace{0cm} \hbox to \hsize{\hfill 
\epsfxsize14cm \epsffile{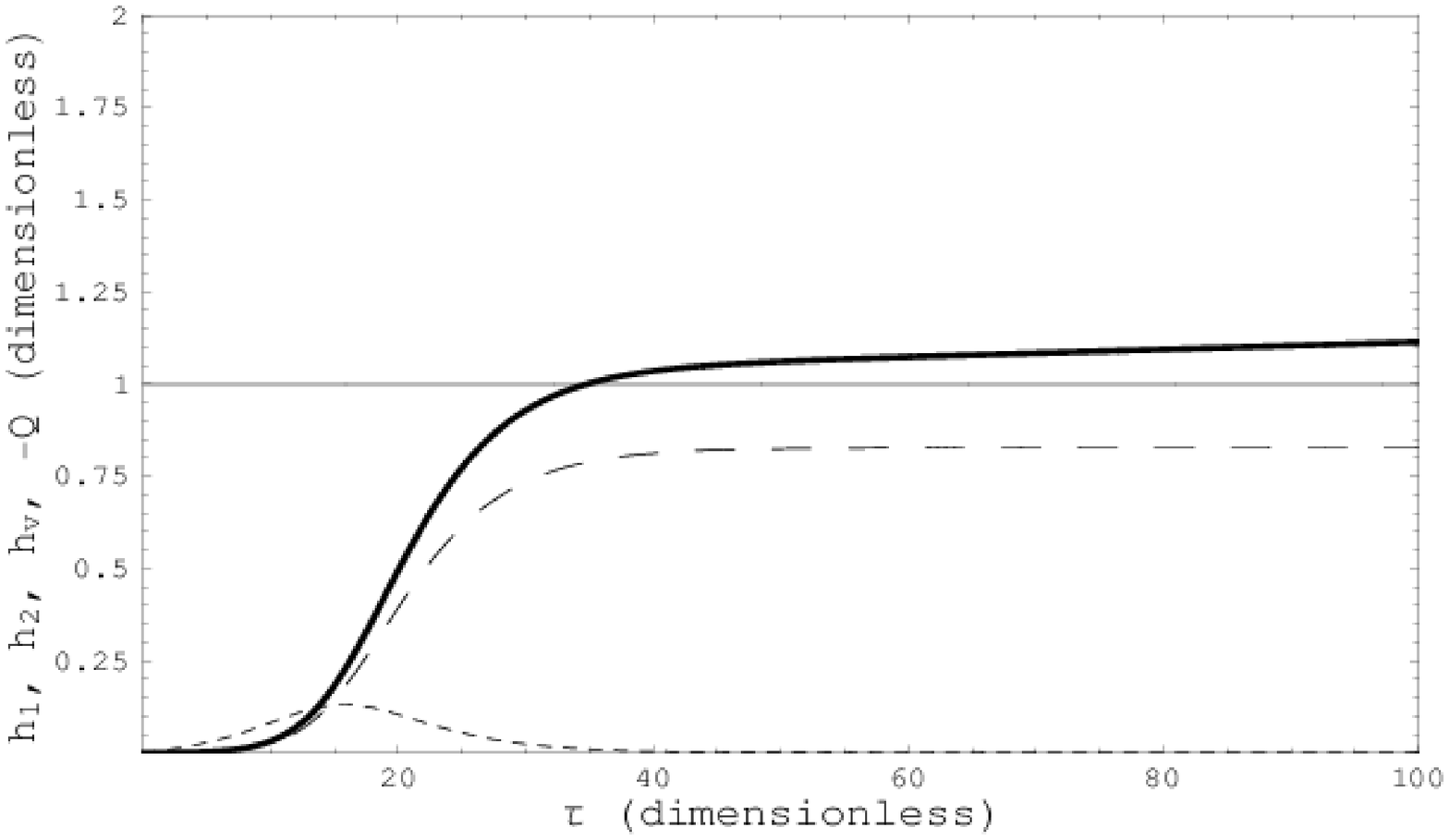} 
\hfill}

\eject

{\bf FIGURE 2cd:} 
\bigskip

\vspace{-.1cm} \hbox to \hsize{ \hfill 
\epsfxsize14cm \epsffile{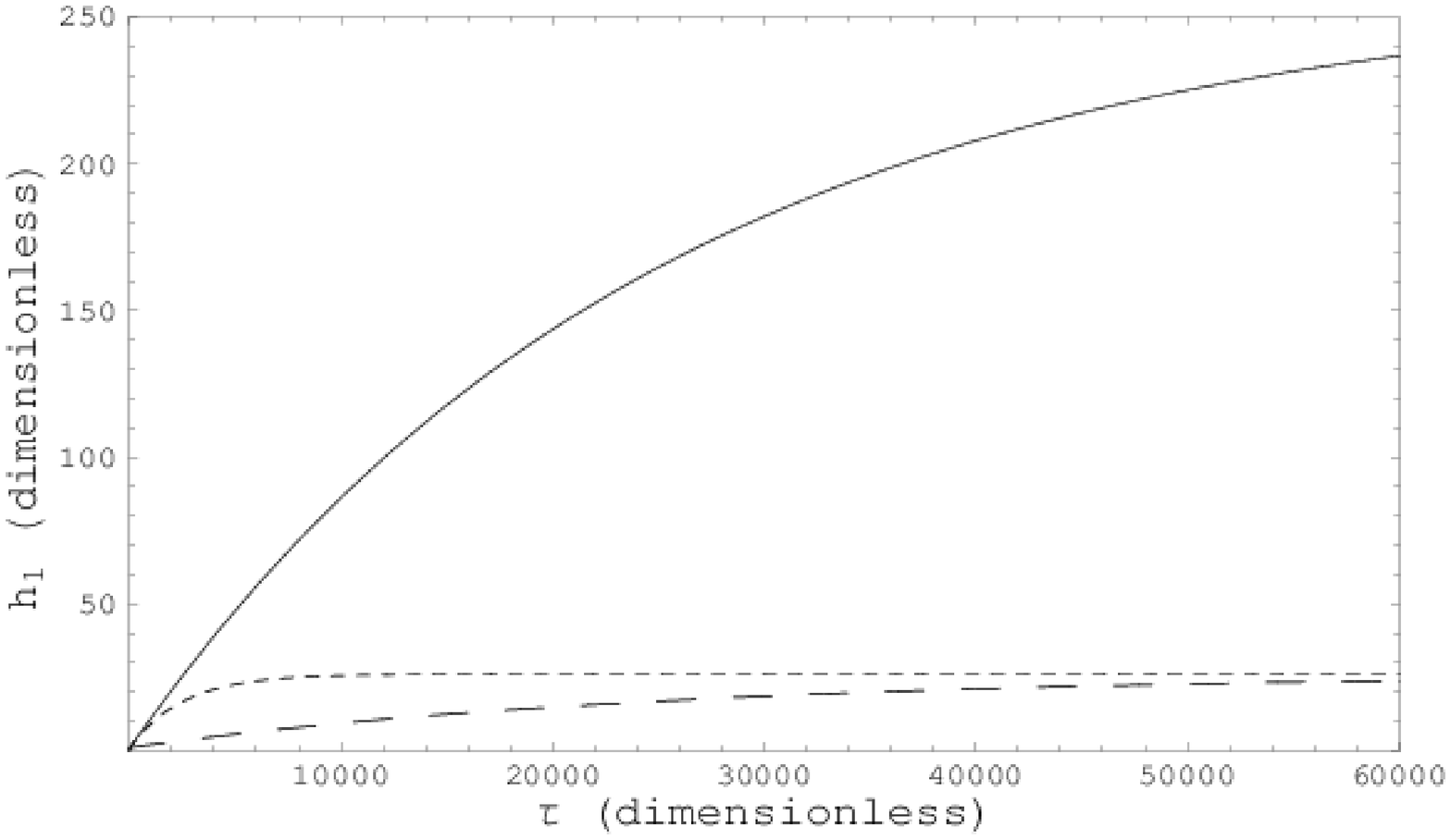} \hfill}
\bigskip
\vspace{-.1cm} \hbox to \hsize{ \hfill 
\epsfxsize14cm \epsffile{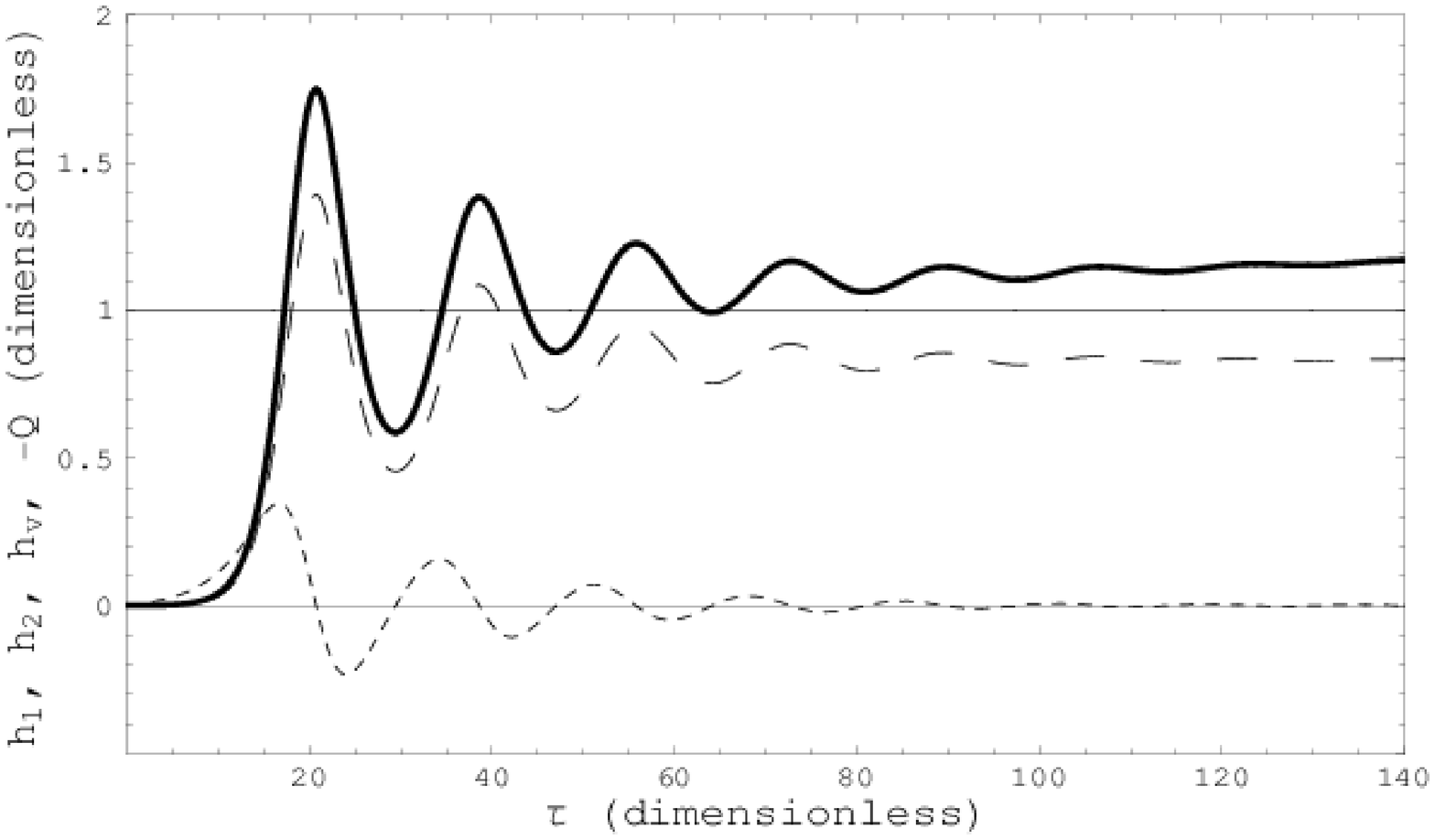} \hfill}



\bigskip

\end{document}